\documentclass[twocolumn,showpacs,preprintnumbers,amsmath,amssymb]{revtex4}
\usepackage{graphicx}% Include figure files
\usepackage{dcolumn}% Align table columns on decimal point
\usepackage{bm}% bold math

\begin{document}

\preprint{APS/123-QED}

\title{
%Unified framework for the charged-defect finite-size supercell
%corrections}
% starting from DFT total energy}
Comparison of charged-defect finite-size supercell correction methods
in a general framework}

\author{Hannu-Pekka Komsa}
 \email{hannu.komsa@tut.fi}
%\author{Ehk{\"{a}} Joku Muu}
%\author{Eero Arola}
\author{Tapio T. Rantala$^1$}
\email{tapio.rantala@tut.fi}
\affiliation{%
Semiconductor Physics Laboratory, Department of Physics,
Tampere University of Technology, FIN-33101 Tampere, Finland
}%

\date{\today}% It is always \today, today,
             %  but any date may be explicitly specified

\begin{abstract}
Starting from the total energy expressions within density functional
theory, we are able to perform a comparison of several currently
used charged-defect finite-size supercell correction schemes
in a unified manner.
This approach also provides a framework for a further development
of corrections not only for DFT supercell calculations,
but also for more advanced methods and for complex geometries.
The comparison is performed for three separate defect cases:
a gallium vacancy in GaAs, a beryllium interstitial in GaAs and
a vacancy in diamond.
We found two methods working sufficiently well for all
three cases: a method which is very similar to one presented by
Freysoldt, \cite{Freysoldt09} and a slightly altered potential
alignment method.
\end{abstract}

%\pacs{71.55.Eq}% PACS, the Physics and Astronomy
                             % Classification Scheme.
%\keywords{Suggested keywords}%Use showkeys class option if keyword
                              %display desired
\maketitle

% Other macros
\newcommand{\react}{$\Rightarrow$}
\newcommand{\mue}[1]{\mu_{\textrm{#1}}}
\newcommand{\mut}[1]{$\mu_{\textrm{#1}}$}
\newcommand{\conc}[1]{$10^{#1}$ cm$^{-3}$}
\newcommand{\conce}[2]{$#1 \times 10^{#2}$ cm$^{-3}$}

\newcommand{\frcoo}[2]{$(\frac{#1}{#2},\frac{#1}{#2},\frac{#1}{#2})$}
\newcommand{\frcoon}[2]{$(-\frac{#1}{#2},-\frac{#1}{#2},-\frac{#1}{#2})$}

\newcommand{\fcite}[1]{\footnote{Ref.\ \onlinecite{#1}}}

\newcommand{\VGa}[1]{V$^{#1}_{\textrm{Ga}}$}
\newcommand{\VC}[1]{V$^{#1}_{\textrm{C}}$}
\newcommand{\BeI}[1]{Be$^{#1}_{\textrm{I}}$}

\newcommand{\tOmega}{\tilde{\Omega}}
\renewcommand{\d}{\text{d}}
\newcommand{\dr}{\d {\mathbf r}^3}

\section{Introduction}

The usual low concentration of defects in connection
with the long-range Coulomb interaction is a difficult
case for the supercell calculations within the density
functional theory (DFT) framework. Especially if
the defect is charged, the calculated formation energies 
strongly depend on the supercell size.
Several correction schemes have been proposed to allow
a calculation of defect properties in an effectively low
concentration by using only a small supercell.
\cite{Makov95, Freysoldt09, VandeWalle04, Schultz00}
However, a consensus on the validity and applicability
of each method seems to be missing, although
numerical comparisons have been done.
\cite{Castleton06}

In this paper,
we start by deriving a generalized approach
for the charged defect supercell calculations,
which is based on the construction and comparison
of the DFT total energy equations for the
supercell and for a much larger cell.
This is useful in several ways:
a) This allows us to properly compare
several contributions of the corrections
and several previous correction methods.
We show how several schemes come out 
as limits or approximations of the general equations.
We particularly concentrate on the method recently introduced
by Freysoldt et al.,\cite{Freysoldt09} and 
also the Makov-Payne \cite{Makov95} and
the potential alignment schemes \cite{VandeWalle04} are reviewed.
b) It shows the required approximations in a clear manner.
Even if the final results look intuitively simple,
the way to get there has a few corners.
c) Paves the way for the development of even better schemes.
We present two more schemes in this paper.
d) It should prove useful in developing correction schemes
for more advanced methods such as hybrid-functionals or $GW$,
and for more complex geometries such as interfaces or clusters.
The first three items are considered in detail in
this article. The last one is briefly considered in
the discussion part, but mostly left for later study.

Finally, we compare these methods for three defects:
gallium vacancy in GaAs, carbon vacancy in diamond and
beryllium interstitial in GaAs.
In order to find the results to compare at,
we use a set of calculations with increasing supercell sizes
to extrapolate the results to the low concentration limit.

\section{Correction scheme}

The defect calculation is usually approached through the concept
of the formation energy, defined as \cite{VandeWalle04}
\begin{eqnarray} \label{eq:forme}
E^{f}[X^q] &=& E_{\rm{tot}}[X^q] -  E_{\rm{tot}}[{\rm{bulk}}] - \nonumber \\
&& \sum_{i}n_i\mu_{i} + q[E_{F} + E_{v} + \Delta V],
\label{eq:formenergy}
\end{eqnarray}
where $E_{\rm{tot}}[X^q]$ is the total energy of the supercell with
the defect $X$ (in the charge state $q$) and $E_{\rm{tot}}[\rm{bulk}]$
is the total energy of the supercell of bare GaAs or GaAsN bulk,
depending on the case.
The chemical potentials $\mu_{i}$ of $n_i$
added or removed atoms allow us to describe various growth
conditions. Here, $E_v$ is the valence band maximum (VBM)
at $\Gamma$-point in the bulk material,
$E_{F}$ is the Fermi energy with respect to $E_v$, and $\Delta V$ is the
shift term used to align the potentials in between the two supercells.

Before moving on to the finite-size supercell considerations, we
divide the comparison of charge defect and bulk cases
in to two parts: comparison of a neutral defect to the bulk, and
the comparison of a charged defect to a neutral defect
\begin{equation}\label{eq:formediv1}
E^f[X^0] = E[X^0] - E[{\rm{bulk}}] - \sum_{i}n_i\mu_{i}
\end{equation}
\begin{equation}\label{eq:formediv2}
E^f[X^q] = E^f[X^0] + \Delta E^f[X^q]
\end{equation}
where
\begin{equation}\label{eq:formediv3}
\Delta E^f[X^q] = E[X^q] - \left[ E[X^0] + q(E_v[X^0] + E_F) \right]
\end{equation}
We see, that in this approach the addition/removal of electrons,
the VBM energy is in the energy reference of the neutral defect.
The implications of this will be discussed later.
Moreover, Comparison of charge densities and potentials between
charged defect and bulk can be difficult, especially in
the case of relaxed geometry.
On the other hand, comparison of charged defect and neutral defect
properties is often much easier.

Naturally, defect calculations within DFT-LDA also suffer
from the band-gap problem. This problem is important to acknowledge
in the analysis of the results, but does not affect the analysis
of the supercell-size correction methods.

\subsection{Charged-defect finite-size supercell corrections}

We will start by writing and comparing the total energies
for neutral defect, charged defect in a small supercell, and 
a charged defect in a much larger supercell.
For the neutral defect charge distribution $\rho_0$
the total energy within the DFT framework is
\begin{equation}
E[\rho_0] = T[\rho_0] + E_{\rm es}[\rho_0] + E_{\rm xc}[\rho_0]
\end{equation}
Adding a localized charge distribution $\delta q$ results
in a redistribution of the surrounding electrons.
We circumvent this problem, by just writing
the total density as $\rho_0 + q_d$.
%In the charged case, the total density is written
%as $\rho_0 + q_d$.
Moreover, we denote the periodically
repeating charge distribution of small supercell as $\tilde{q}_d$.
The total energies for the supercell of volume $\tOmega$ and
a large cell of volume $\Omega$ are then
\begin{eqnarray*}
E^{\tOmega}[\rho_0+\tilde{q}_d] &=& T^{\tOmega}[\rho_0+\tilde{q}_d] +
  E_{\rm es}^{\tOmega}[\rho_0+\tilde{q}_d] +
  E_{\rm xc}^{\tOmega}[\rho_0+\tilde{q}_d] \\
E^{\Omega}[\rho_0+q_d] &=& T^{\Omega}[\rho_0+q_d] +
  E_{\rm es}^{\Omega}[\rho_0+q_d] + E_{\rm xc}^{\Omega}[\rho_0+q_d]
\end{eqnarray*}
Usually, one would like to calculate the formation
energy of a single defect in an infinite crystal, or at least
of much lower concentation of defects than is possible to obtain
in the supercell calculations i.e.,
we would like to have $E^{\Omega}[\rho_0+q_d]$.
Instead, what we get from the supercell calculations is
$E^{\tOmega}[\rho_0+\tilde{q}_d]$.
Thus, what we are looking for in here, is a correction $\Delta E$,
such that $E^{\Omega}[\rho_0+q_d] - E^{\Omega}[\rho_0]$
required for Eq. \ref{eq:formediv3} can be
obtained from $E^{\tOmega}[X^q] - E^{\tOmega}[X^0] + \Delta E$.
%$E^{\tOmega}[\rho_0] = M E^{\Omega}$ and same goes for all
%the contributions of total energy.
Also notice, in the following, we have taken the ionic
configuration to be the same in all three cases.

Due to the locality of the kinetic energy $T$ and % of noninteracting system
the exchange-correlation energy $E_{\rm xc}$, 
and the localization of $q_d$ in the supercell,
we will now assume that
\begin{equation}
T^{\Omega}[\rho_0+q_d] = T^{\tOmega}[\rho_0+q_d] + 
T^{\Omega}[\rho_0]-T^{\tOmega}[\rho_0]
%T^{\Omega-\tOmega}[\rho_0]
\end{equation}
and similarly for $E_{\rm xc}$,
%Since $T^{\Omega-\tOmega} = T^{\Omega}-T^{\tOmega}$, 
so it follows that
$T^{\Omega}[\rho_0+q_d]-T^{\Omega}[\rho_0] =
T^{\tOmega}[\rho_0+q_d]-T^{\tOmega}[\rho_0]$.
Unfortunately, this can not be done for the electrostatic energy, but
we can now write the total energy differences in the large and small
supercell, and find, that their difference, which is the correction
that we are looking for, depends only on the electrostatic energy
difference
\begin{eqnarray*}
\Delta E &=&
E^{\Omega}[\rho_0+q_d] - E^{\Omega}[\rho_0] -
\left( E^{\tOmega}[\rho_0+\tilde{q}_d] - E^{\tOmega}[\rho_0] \right) \\
&=& E_{\rm es}^{\Omega}[\rho_0+q_d] - E_{\rm es}^{\Omega}[\rho_0] -
  \left( E_{\rm es}^{\tOmega}[\rho_0+\tilde{q}_d] -
  E_{\rm es}^{\tOmega}[\rho_0] \right)
\end{eqnarray*}

The electrostatic potential corresponding to $\rho_0$ is denoted as
$V_0$ (consisting of the external potential and the Hartree potential).
%In DFT, the contribution to total energy due to
%electrostatic potential $V_0 = V_{\rm ext} + V_H$ is
The electrostatic energy is (omitting $\dr$ for brevity)
\begin{equation}
E_{\rm es}[\rho_0]
  = \int \rho_0 V_{\rm ext} + \frac{1}{2} \int \rho_0 V_H
  = \int \rho_0 (V_{\rm ext} + \frac{1}{2} V_H)
\end{equation}
where the $\frac{1}{2}$ takes care of the double counting
of electron-electron interactions.
Moving on to the charged case, there is a change
in the Hartree potential $V_H \to V_H + V_{q/0}$ and subsequently
in the electrostatic potential $V_0 \to V_0 + V_{q/0}$.
%(following the notation in Ref.\ \onlinecite{Freysoldt09})
The electrostatic energy is then
\begin{eqnarray}
E_{\rm es}^{\Omega}[\rho_0+q_d] &=& \int_{\Omega} (\rho_0 + q_d + n)
  (V_{\rm ext} + \frac{1}{2}V_H + \frac{1}{2}V_{q/0} )
\end{eqnarray}
where $n = -q/\Omega$ is the neutralizing background.
%These integrals are basically over all crystal.
$\tilde{V}_{q/0}$ is the change of electrostatic potential in
small supercell, which is also, due to the linearity of Poisson
equation, the solution for $q_d$.
In the periodic case, we have similarly added charge
$\tilde{q}_d \approx q_d$, but per unit cell of volume $\tilde{\Omega}$,
so the compensating background charge is $\tilde{n} = -q/\tilde{\Omega}$,
giving
\begin{eqnarray}
E_{\rm es}^{\tOmega}[\rho_0+\tilde{q}_d]
  &=& \int_{\tilde{\Omega}} (\rho_0 + \tilde{q}_d + \tilde{n}) 
   (V_{\rm ext} + \frac{1}{2}V_H + \frac{1}{2}\tilde{V}_{q/0}) 
%  &=& \int_{\tilde{\Omega}} (\rho_0 + \tilde{q}_d + \tilde{n}) \tilde{V}_q  \\
%  &=& \int (q_d + \tilde{n})(\tilde{V}_{q/0}-V_0) + \int \rho_0 \tilde{V}_q
\end{eqnarray}
Notice, that often $\tilde{n}$ (and $n$) term is not included
in DFT codes in the calculation of the electrostatic energy.
However, for the moment, we still leave this term in.

It can be easily shown that 
\begin{equation}
 \int_{\Omega} (q_d + n)
  (V_{\rm ext} + \frac{1}{2}V_H)
 = \int_{\tilde{\Omega}} (\tilde{q}_d + \tilde{n}) 
   (V_{\rm ext} + \frac{1}{2}V_H)
\end{equation}
and
\begin{equation}
\int q_d V_0 = \int \rho_0 V_{q/0}
\end{equation}
%and the $\int \rho_0 V_0$ term cancels, 
so that we are left with
%\begin{equation}\label{eq:corrwrho0}
%\Delta E = 
%\frac{1}{2}\int_{\Omega} (\rho_0 + q_d + n) V_{q/0} -
%\frac{1}{2}\int_{\tilde{\Omega}} (\rho_0 + \tilde{q}_d + \tilde{n}) 
%  \tilde{V}_{q/0}
%\end{equation}
\begin{equation} \label{eq:mainresult}
\Delta E
  = \frac{1}{2} \int_{\Omega} (q_d + n) V_{q/0}
  - \frac{1}{2} \int_{\tilde{\Omega}} (\tilde{q}_d + \tilde{n})\tilde{V}_{q/0}
\end{equation}

We have arrived at a rather intuitive form, which is often taken as
a starting point for developing the defect correction formulae.
Deriving this formula, $q_d$ was defined as the charge difference
between the charged and neutral defect calculations. 
In order to find more simple correction formula, an analytic form
is assumed for $q_d$. In this case, however, one should also
keep in mind that the corresponding potential needs to be
correctly screened, which depends on the whole system.
One can use the static dielectric constant and arrive
at the Makov-Payne correction \cite{Makov95}.
Alternatively, as was done in Ref.\ \onlinecite{Freysoldt09},
$V_{q/0}$ can be divided in to the long- and short-range parts,
where the screening only in the long-range potential is
handled with the static dielectric constant.

%($q_d$ could be calculated
%either from the KS-orbital $|\psi_i|^2$ or
%just taken to be an analytic function such as a delta-function or
%a Gaussian distribution.)
%$\tilde{V}_{q/0}$ can also be extracted from the calculations.
%Unfortunately, we don't know $V_{q/0}$ yet.
%It's not a direct solution of the Poisson equation for $q_d$
%in volume $\Omega$ (with a static dielectric constant),
%since this ignores the microscopic screening effects.
%One can make this approximation, and arrive on
%the Makov-Payne approximation \cite{Makov95}, for example.

\subsubsection{Separation of long and short-range potentials}

In Ref.\ \onlinecite{Freysoldt09}, te long and short-range parts
of the electrostatic potentials were separated as
%The charge difference $q_d$
% is written as a sum of
%an analytic part (with charge $q$) and
%a neutral correction part
%\begin{equation}
%q_d \to q_d + \delta q_d
%\end{equation}
%which results in the long and short range potentials
\begin{equation}
V_{q/0} = V_q^{\rm lr} + V_{q/0}^{\rm sr}
\end{equation}
where $V_q^{\rm lr}$ is the potential solved from $q_d$ using
the static dielectric constant.
Similar separation is done for the $\tilde{V}_{q/0}$, which
is known from the calculation, and can then be used to determine
the short-range potentials. Since $q_d$ is localized to the unit cell,
$\int_{\Omega} q_d V_{q/0}$ can be changed to
$\int_{\tilde{\Omega}} q_d V_{q/0}$ and
%$n \int_{\Omega} V_{q/0}$ is easy to evaluate analytically. 
a little rearrangement results in
\begin{equation}
-\Delta E
  = \frac{1}{2}\int_{\tilde{\Omega}} (q_d + \tilde{n})(\tilde{V}_{q/0}-V_{q/0})
  + \frac{1}{2} \tilde{n} \int_{\tilde{\Omega}} V_{q/0} 
  - \frac{1}{2} n \int_{\Omega} V_{q/0}
\end{equation}
Which equals to Eqs. (6) and (7) in Ref.\ \onlinecite{Freysoldt09},
when $n \to 0$, except the second last term has an extra $\frac{1}{2}$.
The minus-sign comes from the difference in the definition of the potential.
Note also, that when given in this form,
it is possible to calculate the correction
corresponding to any concentration of defects
(i.e., per volume $\Omega$).

In calculations, the averages of potentials
$V_0$, $\tilde{V}_q$, and $\tilde{V}_{q/0}$ are all set to zero.
%We also had the condition of Eq. \ref{eq:potshiftdif}. This
%means that we have to lose the condition that potential $V_{q/0}$
%would approach zero at long distances from the defect.
%Instead $V_{q/0} \to D$, as $r \to \infty$.
Next, we can choose $V_{q/0}$ such that it approaches zero as $r \to \infty$.
% and choose the shifts in Eq. \ref{eq:rhopotcond} accordingly.
Moreover, let's write
$\tilde{V}_{q/0}^{\rm sr} = \sum_R V_{q/0}^{\rm sr} + C$ 
as in Ref. \cite{Freysoldt09}.
With these considerations
\begin{eqnarray}
-\Delta E
  &=& \frac{1}{2} \int_{\tilde{\Omega}} (q_d + \delta q_d + \tilde{n})
    (\tilde{V}_q^{\rm lr}-V_q^{\rm lr} + C) \nonumber \\
  &+& \frac{1}{2} \tilde{n} \int_{\tilde{\Omega}} (V_q^{\rm lr} + V_q^{\rm sr})
  - \frac{1}{2} n \int_{\Omega} (V_q^{\rm lr} + V_q^{\rm sr})
\end{eqnarray}
since
\begin{equation}
\int V_q^{\rm sr} = \int \tilde{V}_q^{\rm sr} - C
  = \int \tilde{V}_{q/0} - \tilde{V}_q^{\rm lr} - C
  = \int C
\end{equation}
and $\tilde{n}\int_{\tilde{\Omega}} C = n\int_{\Omega} C$, we get
\begin{eqnarray}
-\Delta E
  &=& \frac{1}{2} \int_{\tilde{\Omega}} (q_d + \delta q_d + \tilde{n})
    (\tilde{V}_q^{\rm lr}-V_q^{\rm lr}) \nonumber \\
  &+& \frac{1}{2} \tilde{n} \int_{\tilde{\Omega}} (V_q^{\rm lr} + C)
   - \frac{1}{2} n \int_{\Omega} (V_q^{\rm lr} + C) \\
  &=& \frac{1}{2} \int_{\tilde{\Omega}} (q_d + \delta q_d + \tilde{n})
    (\tilde{V}_q^{\rm lr}-V_q^{\rm lr}) \nonumber \\
  &+& \frac{1}{2} \tilde{n} \int_{\tilde{\Omega}} V_q^{\rm lr}
   - \frac{1}{2} n \int_{\Omega} (V_q^{\rm lr}).
\end{eqnarray}

Finally, because $\int_{\tilde{\Omega}} \tilde{n} \tilde{V}_q^{\rm lr} = 0$
and $\frac{1}{2} n \int_{\Omega} V_q^{\rm lr} \to 0$ as
$\Omega \to \infty$, a very simple result is obtained:
\begin{equation} \label{eq:finalcorra}
-\Delta E
  = \frac{1}{2} \int_{\tilde{\Omega}} (q_d + \delta q_d)
    (\tilde{V}_q^{\rm lr}-V_q^{\rm lr})
\end{equation}
We see that there is no dependence on the potential shift
$C$, nor shifts $A$ or $B$, but this is not really
in disagreement with Ref.\ \onlinecite{Freysoldt09}.
In our approach, similar potential alignment term comes from taking
the VBM with respect to the neutral defect energy reference
(see Eqs. \ref{eq:formediv1}--\ref{eq:formediv3}).

%[Or,
%\begin{equation} \label{eq:finalcorrb}
%\tilde{E} - E
%  = \frac{q^2}{2 \varepsilon_r} \int_{\tilde{\Omega}} (q_d' + \delta q_d)
%    (\tilde{V}_1^{\rm lr}-V_1^{\rm lr})
%\end{equation}
%where $q_d'$ is gaussian charge density of $q=1$ with
%the corresponding potentials $V_1$ (linearity of Poisson equation).
%The integral is rather difficult to handle analytically
%(although probably not impossible), but it is easy to calculate
%numerically and the values as a function of width of the charge
%distribution and the unit cell sizes can be plotted or tabulated.]

%[Even if I'm wrong about the $1/2$ difference, Eq. (13)
%of \cite{Freysoldt09} still reduces to a much simpler form:
%\begin{equation}
%-\Delta E
%  = \frac{1}{2} \int_{\tilde{\Omega}} \left[ q_d
%    (\tilde{V}_q^{\rm lr}-V_q^{\rm lr}) + \tilde{n}V_q^{\rm lr} \right]
%\end{equation}
%and $\Delta_{q/0} = qC$]

%Some values using Eq. \ref{eq:finalcorrb} are given in
%figure \ref{fig:corrections}
%table \ref{tab:corrections}

%\begin{table}
%\caption{\label{tab:corrections}
%Tabulated corrections
%}
%\begin{ruledtabular}
%\begin{tabular}{lcc}
%\hline
%\end{tabular}
%\end{ruledtabular}
%\end{table}

\subsubsection{Makov-Payne}

\begin{figure}
\begin{center}
  \includegraphics[width=8.0cm]{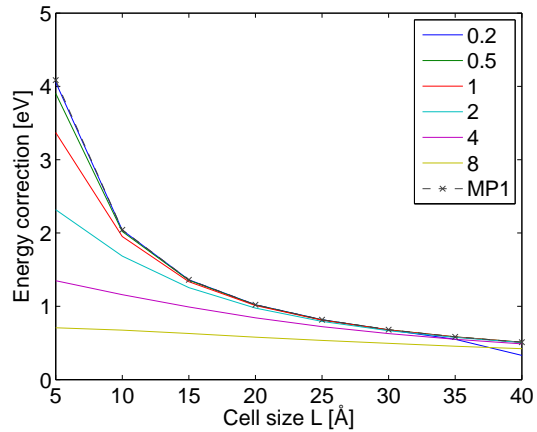}
\end{center}
\caption{\label{fig:corrLandwid}
The correction as a function of the width of the gaussian
charge density.
% (Small bug in the 0.2 result...)
(Color online.)
}
\end{figure}

Makov-Payne correction \cite{Makov95} is
\begin{equation}
\Delta E = q^2 \alpha / 2 \varepsilon_r L +
  \frac{2 \pi q Q_r}{3 \varepsilon_r L^3}
\end{equation}
where
\begin{equation}
Q_r = \int q_d r^2
\end{equation}
is the second radial moment of the density difference.
For SC lattice $\alpha = 2.8373$ \cite{Leslie85}.
When $q_d$ approaches delta-function, the Eq. \ref{eq:finalcorra}
approaches Makov-Payne equation.
This is demonstrated in Figure \ref{fig:corrLandwid}
for different supercell sizes and widths of $q_d$.
Similar calculations were also presented in
Ref.\ \onlinecite{Segev03}.

When $q_d$ is estimated as a gaussian distribution,
the correction becomes somewhat smaller and the dependence on
the supercell size decreases.
Also note, that for a Gaussian $q_d$,
some of these terms can be calculated analytically
(see e.g. Ref.\ \onlinecite{LentoThesis03}.)

\subsection{Potential alignment method}

In the potential alignment scheme, the difference of the electrostatic
potential far from defect with respect to the bulk, $\Delta V$, is read,
essentialy resulting in energy correction $\Delta E = q \Delta V$.

We calculated $\Delta V$ for a simple model charge density
of a (nearly) point charge and a neutralizing background
in a simple-cubic supercell geometry,
expecting to obtain a simple dependence of $\Delta V$
on dielectric constant and supercell size.
%Simply calculating the potential of a nearly point charge,
%in a cubic cell and with $V_{G=0} = 0$,
Indeed, we find the following form for the potential
at the farthest point
\begin{equation}\label{eq:analpotal}
\Delta V \approx 0.78 \frac{q}{\varepsilon_r L}
\end{equation}
where $L$ is given in the units of bohr and the potential
in the units of Hartree. The energy correction has again
the same $q^2/\varepsilon_r L$ scaling although the constant
is smaller than in the Makov-Payne correction for the cubic
cell ($2.8373/2 \approx 1.419$). 
Later in this article, this energy correction
is called analytic potential alignment.

The similarity of the calculated and analytic potential alignments
can be seen in Figures \ref{fig:VGacorr}, \ref{fig:VCcorr}, and
\ref{fig:BeIcorr}.

\subsection{Energy comparison}

Here we consider how to best evaluate
the VBM energy $E_v[X^0]$ in Eq. \ref{eq:formediv3}.
In case there are no defect states near the VBM,
there is no problem in the first place.
However, even if there are defect states mixing with VBM,
in the case of neutral defect, the electrostatic potential should
converge to the bulk value fast (faster than $r^{-1}$), and it
can be obtained easily from the comparison of the potentials
of the neutral defect calculation and of the bulk calculation.

We find that taking the electrostatic potential difference
from the ion cores or the plane-averaged over the supercell
far from defect gives very similar result. Especially so
in the unrelaxed geometries,
but in the relaxed geometries the former probably proves easier
to use. The usage of $E_v[X^0]$ from the neutral defect calculation
instead of bulk $E_v$ is later denoted as VBM alignment.
Notice the difference to the potential alignment method, where
the potential difference is taken for each charged supercell
of interest.

\subsection{Complete method and discussion}

Here we outline two correction schemes that seemed to work
well for the studied cases.

To sum up, the final correction scheme is then the following:
\begin{enumerate}
\item Obtain VBM alignment ($E_v[X^0]$) from the comparison
of neutral defect and bulk.
\item Find gaussian $q_d$ which gives close match to
the resulting change in the electrostatic potential between
charged and neutral defects. (When far from the defect.)
% And especially so, that the potential matches far from defect
% this guarantees that the ``short range part'' goes to zero...
\item Apply Eq. \ref{eq:finalcorra} to get the final formation
energy, along with the Eqs. \ref{eq:formediv1}--\ref{eq:formediv3}.
\end{enumerate}
This is the scheme I.

We also found, that the following scheme (scheme II)
seems to work surprisingly well:
\begin{enumerate}
\item Obtain the VBM alignment ($E_v[X^0]$) as before
\item Correct the formation energies by using the analytical
form for the potential alignment Eq. \ref{eq:analpotal}.
\end{enumerate}

%...Try to find $q_d$ width either from the charge-distribution
%or matching the potentials.

\section{Applications}

In all of the calculations, we use planewave density-functional
theory code {\scshape VASP} within the PAW-LDA formalism.
\cite{kres1,kres2,kres3}
In GaAs calculations, we have chosen gallium to have 3d frozen
in the core and 400 eV cutoff.
In diamond calculations, the cutoff is 500 eV.
We use 64-atom supercells with 
%$2\times2\times2$ and
$4\times4\times4$ $\bf{k}$-points, 216-atom supercells
with $2\times2\times2$ $\bf{k}$-points, and 512-atom supercells with 
%$1\times1\times1$ and 
$2\times2\times2$ $\bf{k}$-points.
%Moreover, we took care that the FFT-mesh was scaled properly.
% Resulted in 80 meV error of pure GaAs energy.

We will now consider three test cases: a gallium vacancy in GaAs,
a vacancy in diamond, and a beryllium interstitial in GaAs.
All of these are calculated in both the unrelaxed and
the relaxed geometries.
%The uncorrected formation energies are collected in
%Figure \ref{fig:forme}.
%In this, and 
In all of the following formation energy figures,
lines of the form $aL^{-1} + bL^{-3} + c$ are fitted to the
calculated values to obtain extrapolated values in
the limit $L^{-1} \to 0$ ($L \to \infty$).

%\begin{figure*}
%\begin{center}
%  \includegraphics[width=5.5cm]{figs/VGaextrap.eps}
%  \includegraphics[width=5.5cm]{figs/BeIextrap.eps}
%  \includegraphics[width=5.5cm]{figs/VCextrap.eps}
%\end{center}
%\caption{\label{fig:forme}
%The formation energies as a function of inverse cell size
%for \VGa{0} and \VGa{-3} (a),
%\BeI{0} and \BeI{+2} (b),
%and \VC{0}, \VC{+2} (c, lower part) and \VC{-4} (c, upper part).
%Solid lines are with denser k-point sets and dashed
%lines with coarser k-point sets.
%``u'' is for unrelaxed and ``r'' is for relaxed.
%(Color online.)
%}
%\end{figure*}

\subsection{Gallium vacancy in GaAs}

\begin{figure*}
\begin{center}
  \includegraphics[width=8.0cm]{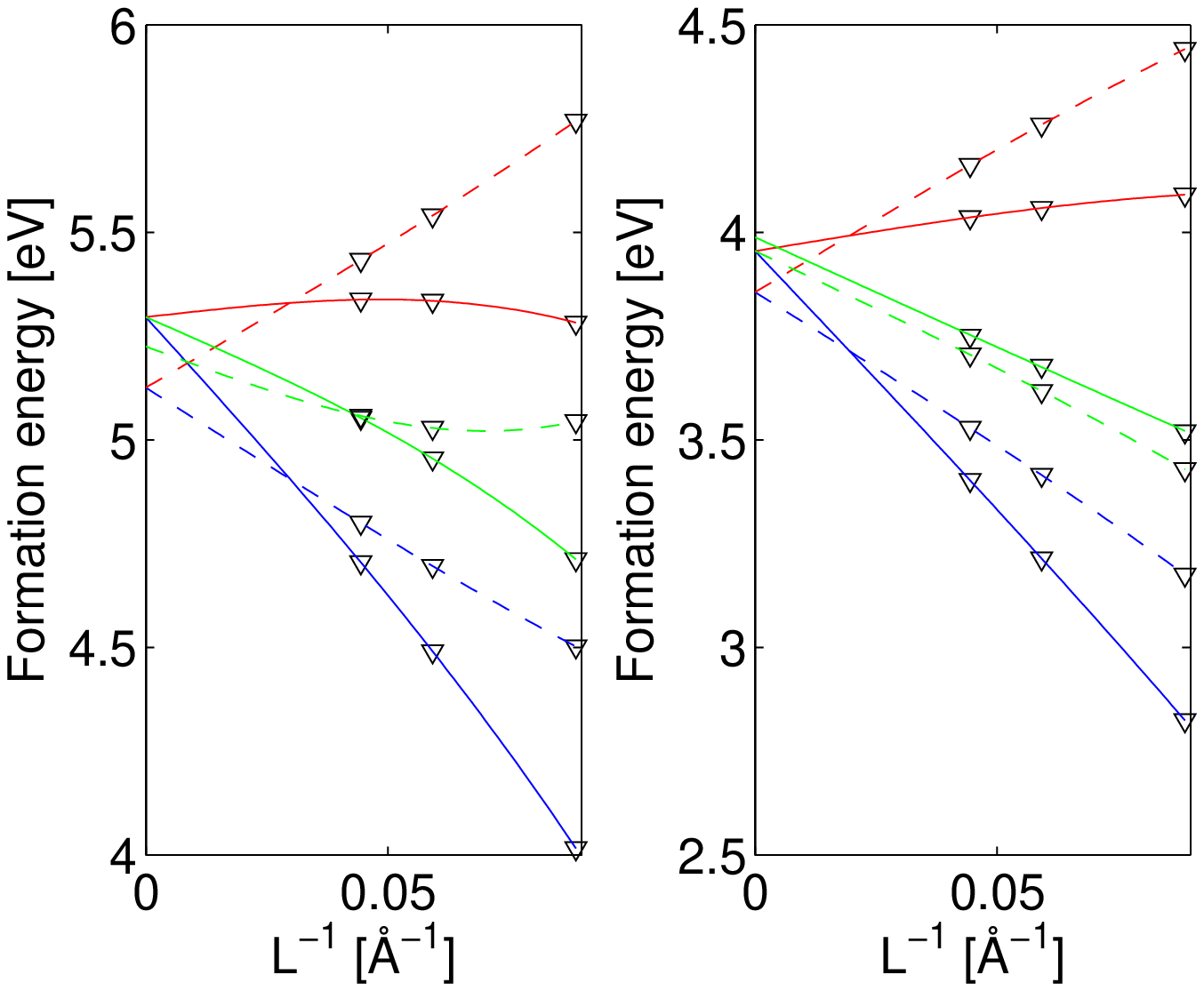}
  \includegraphics[width=8.0cm]{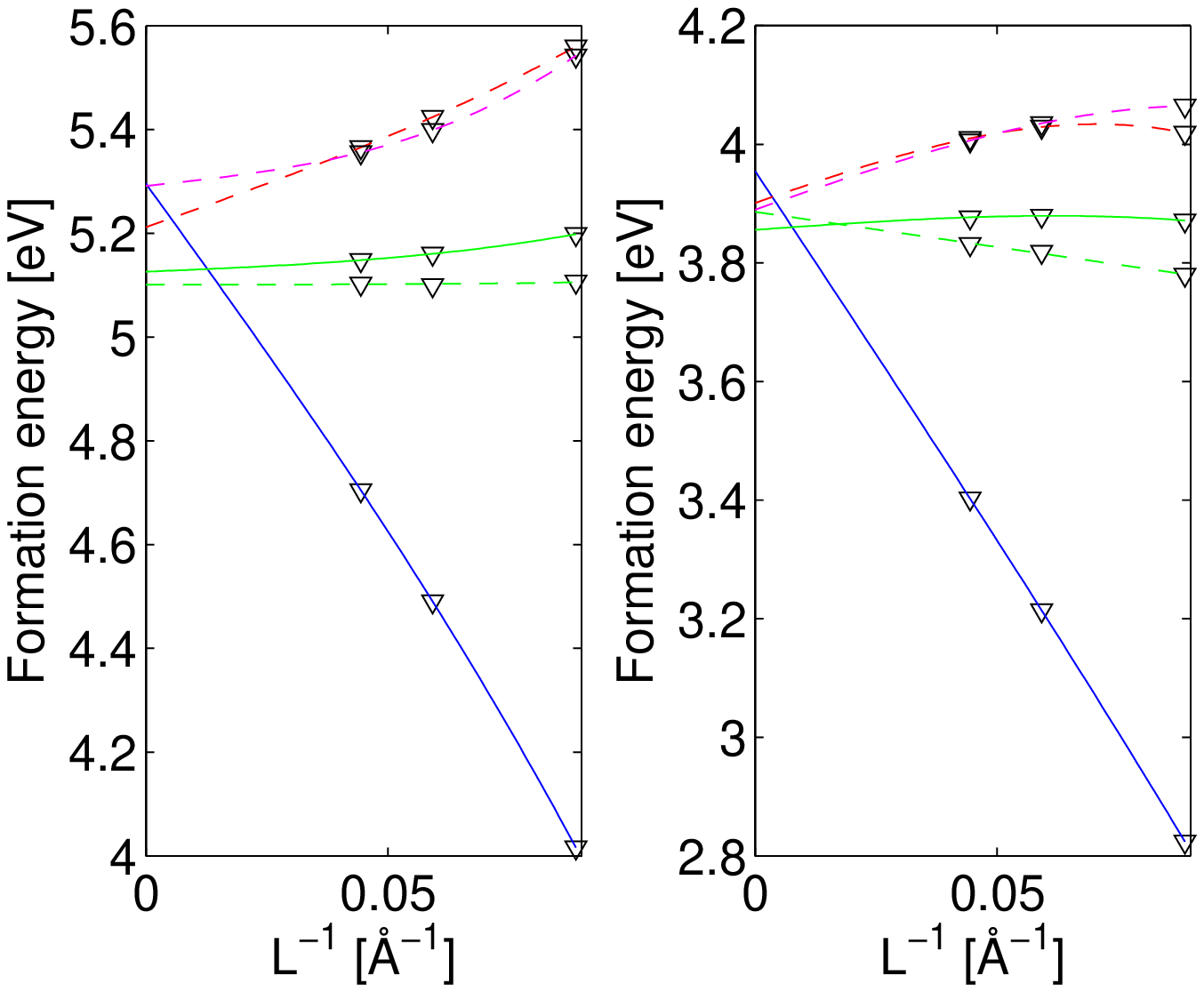}
\end{center}
\caption{\label{fig:VGacorr}
Corrections for \VGa{-3} in unrelaxed (left subfigure)
and relaxed (right subfigure) cases.
Color coding in left figure:
no corrections (blue, solid),
VBM alignment (blue, dashed),
Madelung correction (red, solid),
Madelung with VBM alignment (red, dashed),
analytic potential alignment (green, solid), and
calcuted potential alignment (green, dashed).
%and Freysoldt (magenta).
Color coding in right figure:
no corrections (blue, solid),
analytic potential alignment with VBM alignment (scheme II) (green, solid),
calculated potential alignment with VBM alignment (green, dashed),
gaussian charge correction with VBM alignment (scheme I) (red, dashed)
gaussian charge correction with charged cell VBM alignment (magenta, dashed).
(Color online.)
}
\end{figure*}

The application of several correction schemes for
the gallium vacancy are shown in Figure \ref{fig:VGacorr}.
Makov-Payne scheme seems to work well, potential alignment
underestimates the formation energies.
Scheme I tends to overestimate the formation energies
about 100 meV in large supercells and somewhat more
in the 64-atom supercell.
With VBM alignment, formation energies converge to a lower
value. Consequently, scheme II energies are slightly lower
than the extrapolated value from the uncorrected energies.
However, the energies have very little variation over
the supercell sizes in this scheme.
%The correction by Freysoldt et al. seems indeed to be
%missing the factor $1/2$.
%It is unclear, if it is well justified to perform
%extrapolation on the corrected values.

%The potentials $\tilde{V}_q^{\rm lr}$ and $V_q^{\rm lr}$
%and the calculated potential are shown in
%Figure \ref{fig:VGaVCpots}.

%On the other hand,
%if the extrapolation is not needed, then the whole k-point
%convergence issue becomes less important.

%[Band edge occupations occur, which
%reflect in to $L^{-1}$ scaling \cite{Castleton04}.]

%\begin{figure}
%\begin{center}
%  \includegraphics[width=8.0cm]{figs/VGachgdifr.eps}
%\end{center}
%\caption{\label{fig:VGachgdifr}
%The charge difference of \VGa{-3} and the bulk as a function
%of distance to the defect. The average is shown as a solid line
%and in the inset the averages for \VGa{0} (converging to -3)
%and \VGa{-3} (converging to 0) are compared.
%(Color online.)
%}
%\end{figure}

\subsection{Carbon vacancy in diamond}

Similar to Ref.\ \onlinecite{Freysoldt09}, we calculate
the carbon vacancy in diamond for neutral, $+2$ and $-4$
charge states. In this case, due to the large band gap of
diamond, there is no energy overlap of the defect states and
the band edge states. This, along with large variations
in stable charge states, makes the case particularly
suitable for studying finite-size supercell interactions
without having to worry about the band-gap errors.

\begin{figure*}
\begin{center}
  \includegraphics[width=8.0cm]{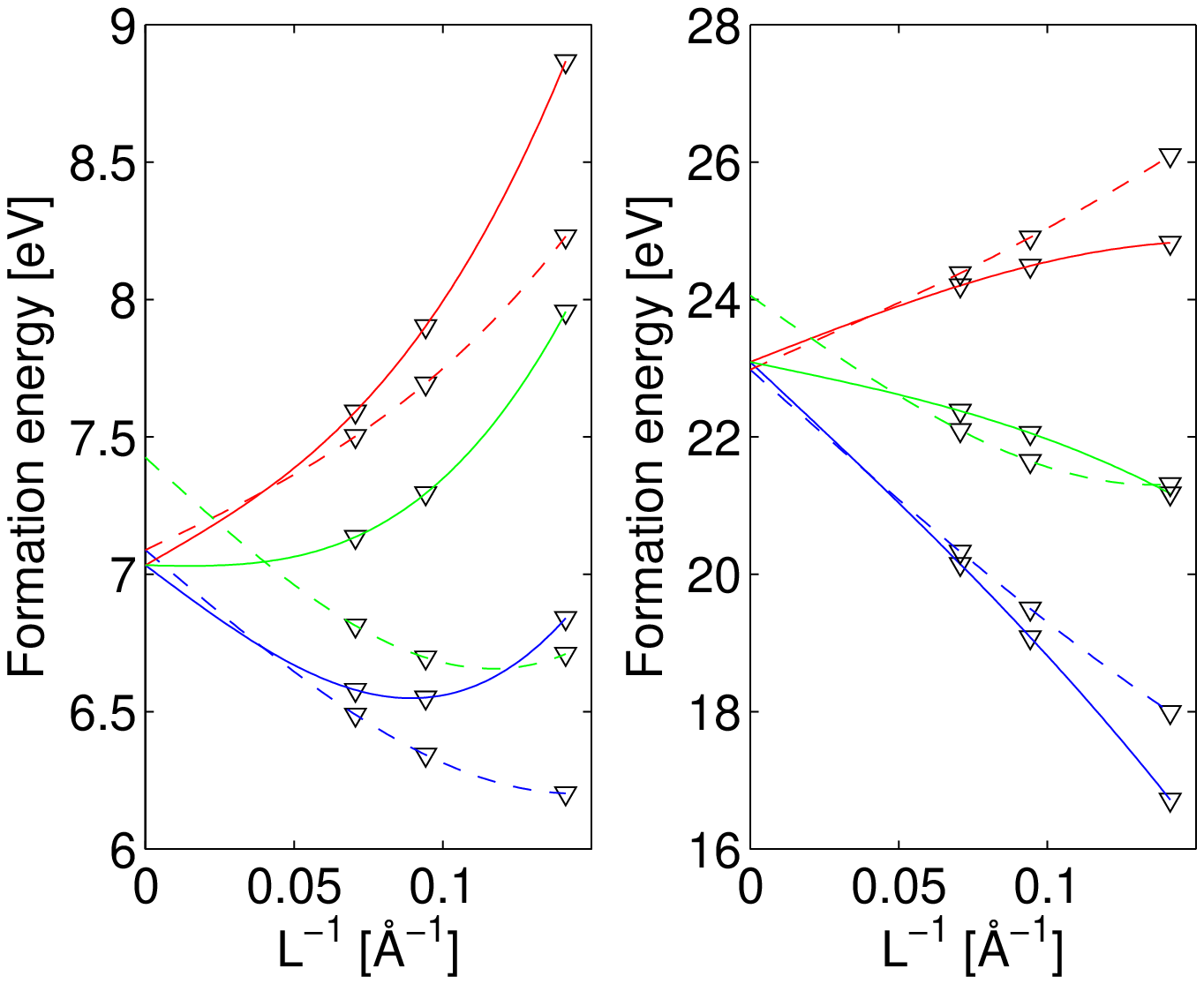}
  \includegraphics[width=8.0cm]{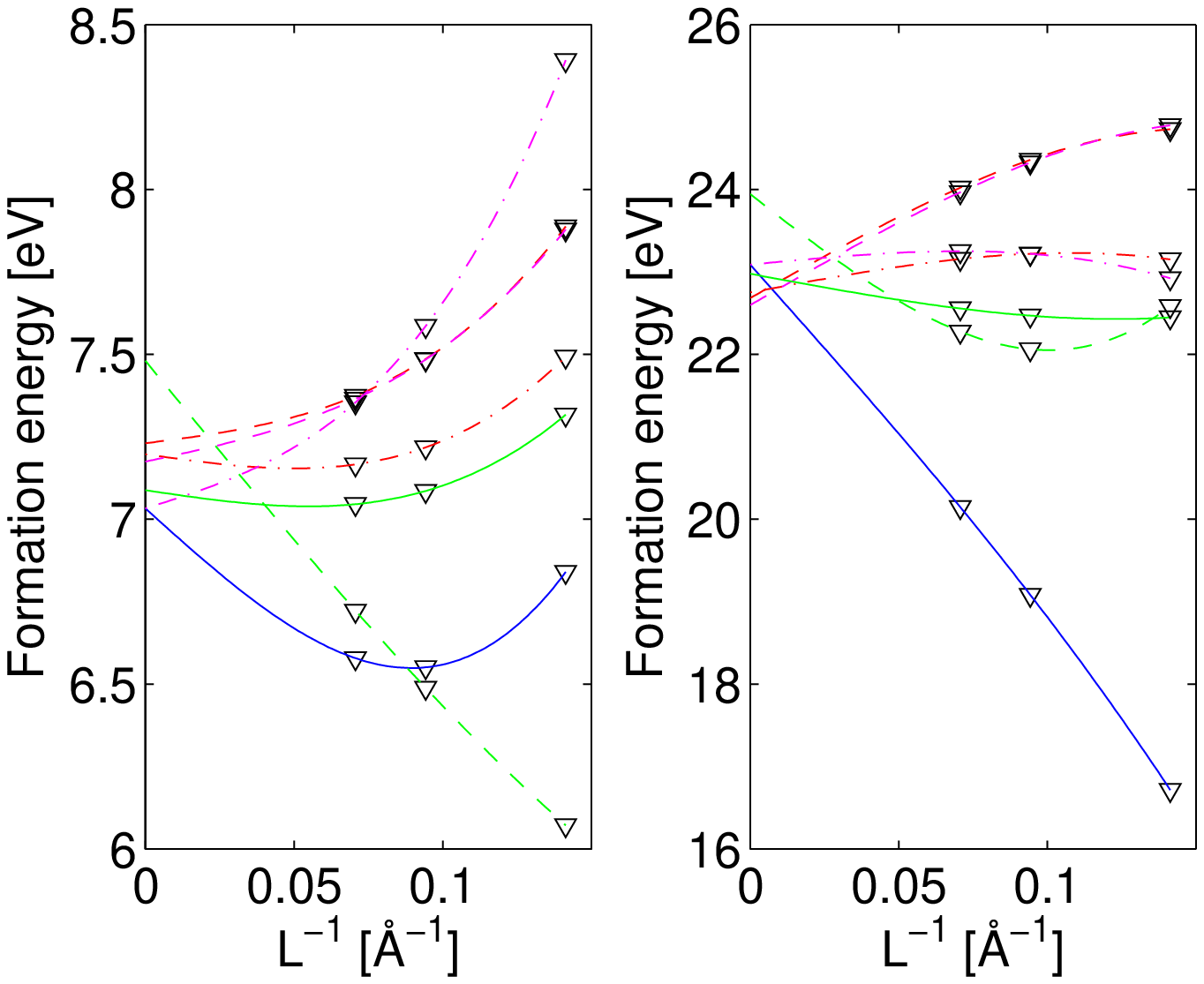}
\end{center}
\caption{\label{fig:VCcorr}
Corrections for unrelaxed \VC{+2} (left subfigure)
and unrelaxed \VC{-4} (right subfigure) cases.
Color coding as in Figure \ref{fig:VGacorr},
except for the dash-dotted lines:
Makov-Payne with effective charges (magenta, dash-dot),
scheme I with effective charges (red, dash-dot).
(Color online.)
}
\end{figure*}

The formation energies are shown in figure \ref{fig:VCcorr}.
Especially for the $+2$ case, Makov-Payne notably overcorrects.
Potential alignment seem to work fairly well.
Scheme II surprises again, with very little variation and energies
close to the extrapolated value.
Note, however, that for the \VC{+2} defect the analytic and
calculated potential alignments differ considerably.

\begin{figure}
\begin{center}
  \includegraphics[width=8.0cm]{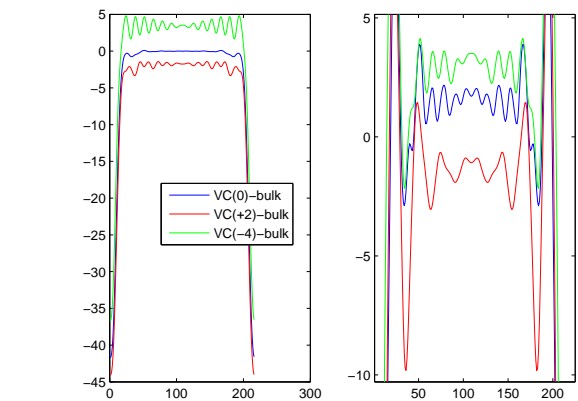}
\end{center}
\caption{\label{fig:VCchgdifr}
The xy-averaged charge difference
(multiplied with cell volume) of \VC{+2} and \VC{-4}
to the bulk. Unrelaxed geometry (left) and
relaxed geometry (right)
%, and a magnification of the unrelaxed geometry (down)
with horizontal lines roughly
denoting the charge difference far from defect.
(Color online.)
}
\end{figure}

Scheme I overcorrects again, but more than in the case of
gallium vacancies.
A reason can be traced to the charge distribution difference
among the charged and neutral cases.
The charge distribution difference of the relaxed \VC{+2} and
\VC{-4} with respect to the bulk case are shown in
Figure \ref{fig:VCchgdifr}.
The charge difference seems good in the neutral case,
but in the charged cases, the added/removed charge is almost
completely delocalized, except that e.g. \VC{-4} does not
converge to $-4$, but $-3.5$. As the Poisson equation solution
is also governed by the charge distribution away
from the defect (large volume at large distance $r$),
it seems worth trying to use these charges in
the correction schemes.
This fixes nicely the formation energies produced by scheme I,
and also improves Makov-Payne results, except for the
unrelaxed 64-atom supercell result.

\begin{figure}
\begin{center}
  \includegraphics[width=8.0cm]{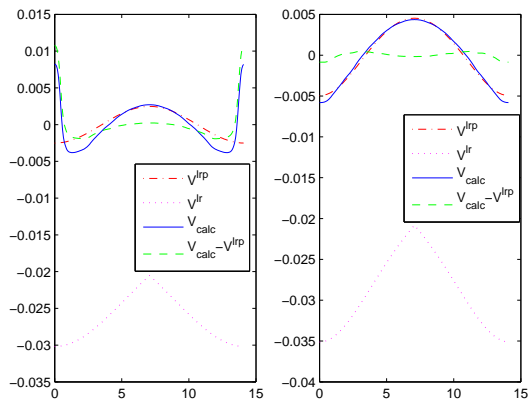}
\end{center}
\caption{\label{fig:VGaVCpots}
The xy-averaged potentials from 512-atom supercell,
similar to Fig. 2 in Ref.\ \onlinecite{Freysoldt09},
%Up: \VGa{-3} against bulk.
%Down: 
of \VC{+2} against bulk (left) and against \VC{0} (right).
Blue solid: from calculation. Red dash-dotted: analytic periodic.
Green dashed: the difference. Magenta dotted: analytic aperiodic.
(Color online.)
}
\end{figure}

In any case, these charge distribution graphs
seem to contradict with the basic premise of
the correction methods that there is a localized charge.
Still, the corrections work fairly well.
The occupied/emptied state was localized around the defect
(not a host band),
meaning that there must be a compensating change
in the valence band electron density.
This behavior in the charge differences
was found in all three defect cases.
%Is the localized charge now only in the ion?

The potentials $\tilde{V}_q^{\rm lr}$ and $V_q^{\rm lr}$
and the calculated potential are shown in
Figure \ref{fig:VGaVCpots}.
We see that the potentials match very well when compared
to the neutral case. This justifies the division of
formation energy calculation into two parts.

%This would be also suggested by the potentials
%in Figure \ref{fig:potlocr}. 
%Interestingly, the potentials in the neutral and charged cases
%behave in a very similar manner, as if there was a localized
%charge at the defect in both cases. Why then, are the formation
%energies so different for the neutral and charged cases!!!???
%This is most likely due to the CB or VB occupations.
%On the other hand, in the neutral case, the charge difference
%to the bulk case shows a delocalized charge density
%integrating to $-3e$.
%This is due to the incorrect occupation of the delocalized
%host states. 

\subsection{Beryllium interstitial in GaAs}

\begin{figure*}
\begin{center}
  \includegraphics[width=8.0cm]{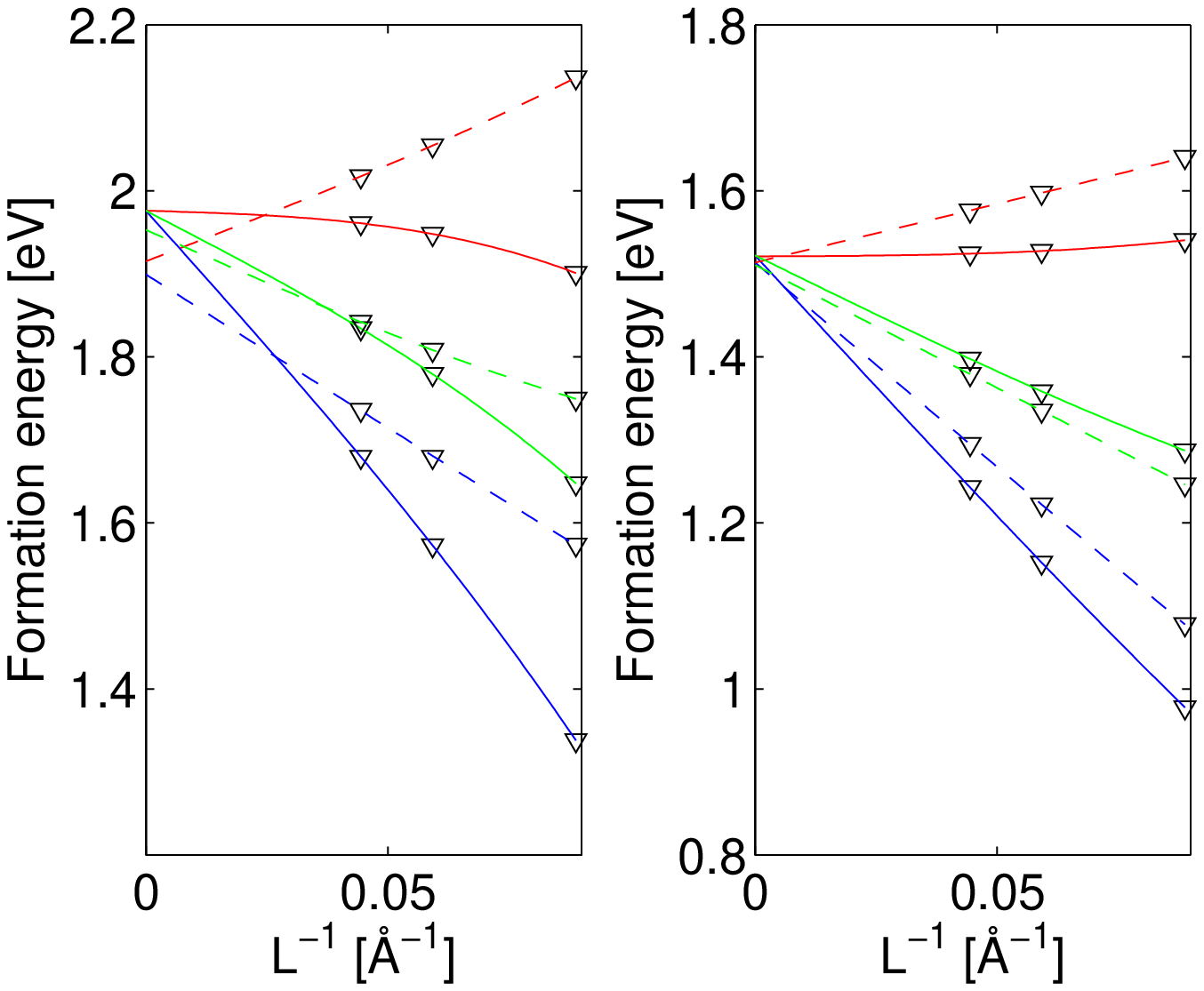}
  \includegraphics[width=8.0cm]{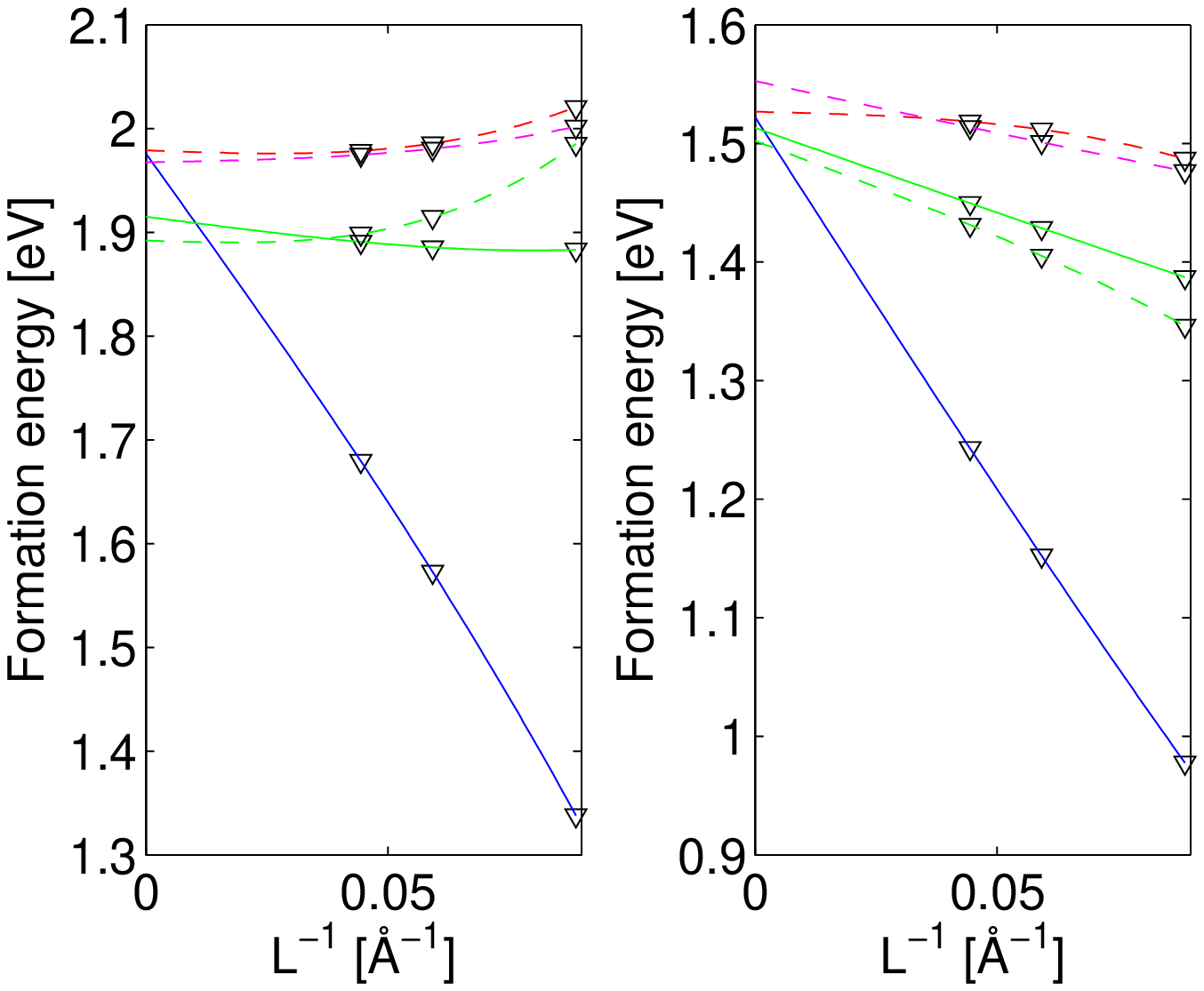}
\end{center}
\caption{\label{fig:BeIcorr}
Corrections for \BeI{+2} in unrelaxed (left subfigure)
and relaxed (right subfigure) cases.
Color coding as in Figure \ref{fig:VGacorr}.
(Color online.)
}
\end{figure*}

In order to test the method with something other than a vacancy,
we calculated beryllium interstitial in GaAs. It was
chosen, because we know from our previous studies \cite{Komsa09},
that the defect states are well localized with very little
geometric distortion.
The formation energies are shown in Figure \ref{fig:BeIcorr}.
General features for the neutral and charged cases are
similar to the gallium vacancy case.

Makov-Payne corrections work even better than
in the case of gallium vacancy. Potential alignments without
VBM correction tend to somewhat underestimate formation energies
as before. Here, scheme I works really well, and scheme II also
relatively well even if it extrapolates again to a lower value.

It seems, that using only the calculated potential alignment
already corrects the formation energies about halfway,
giving a fair confidence on the results reported in our
previous study of beryllium defects.\cite{Komsa09}

Comparison of the relaxed and unrelaxed cases for all defect types
(and as can be seen in Figures \ref{fig:VGacorr} and \ref{fig:BeIcorr})
show no systematic difference in the applicability of the correction methods.
During the derivation, these effects were approximated to be
small, and this would indeed seem to be the case, although
these defect cases were chosen especially chosen to show no
major relaxation effects.

%\begin{figure*}
%\begin{center}
%  \includegraphics[width=8.0cm]{figs/VGapotlocr.eps}
%  \includegraphics[width=8.0cm]{figs/BeIpotlocr.eps}
%\end{center}
%\caption{\label{fig:potlocr}
%Potentials at the Ga cores (left) and As cores (right)
%for \VGa{} and \BeI{}.
%(Color online.)
%}
%\end{figure*}

%\begin{figure}
%\begin{center}
%  \includegraphics[width=8.0cm]{figs/BeIchgdif1.eps}
%\end{center}
%\caption{\label{fig:chgdifs}
%Charge difference of \BeI{+2} to \BeI{0}.
%(Color online.)
%}
%\end{figure}

\section{Discussion}

%In Figure \ref{fig:forme} is also shown the comparison
%of formation energies calculated with low k-point density and
%high k-point density. The changes in formation energies are
%``only'' up to 200 meV, but this reflects in to huge changes 
%in the extrapolated formation energies.
%Therefore, consistent choice of k-point sets is required
%for the extrapolation.
Another problem with the extrapolation method is the amplification
of errors in the set of calculations. If one of the calculations has
an error of 100 meV for any reason, it can result in 1 eV difference in
the extrapolated formation energy.
%Then again, if the extrapolation amplifies the errors dramatically,
Thus, it is advantageous to get rid of the extrapolation scheme.
Taking this idea even further, 
a working correction scheme might allow to use lower precision
in the calculation. One possibility would be 
to use coarser k-point meshes, although naturally other
properties might also degrade.
For example, it has been reported that coarse k-point mesh
can lead to incorrect geometries.
\cite{Puska98, Shim05,Castleton04} % Check Shim05 and references therein.
%[Note the discussions in \cite{Castleton04,Shim05}]
%In the cases studied here, however, at least the charge densities
%in the two cases are very similar.

Extending these correction schemes to more complex geometries such
as interfaces requires again a proper model for the screening.
This is straightforward enough. Alternatively, one could try to
divide the space even further and write the total energy as the
sum of the contributions in these regions.
So far, very limited number of studies has been performed on
defects at interfaces and warrants further investigations.

When developing a correction for e.g the $GW$ method,
to the first approximation, the same trick could be done for the
self-energy as was done here for the XC-energy, eventually yielding
the same corrections. Of course, $GW$ provides a proper
dielectric function for the screening which could be taken
advantage of.
Unfortunately, as $GW$ calculations usually do not provide total
energies,
a somewhat different approach is probably needed.
%it might be better to start from the total energy equation
%with the sum of eigenenergies,
%\begin{equation}
%E[\rho_0] = \sum_i \varepsilon_i - \frac{1}{2} \int V_H \rho_0 +
%  E_{\rm xc}[\rho_0] - \int V_{\rm xc} \rho_0
%\end{equation}
%although then the procedure is probably quite different to
%what is presented in this article.

\section{Conclusions}

A comparison of several previously introduced defect correction
methods and a few new ones are compared in both analytical and
numerical levels. First, we divide the problem of comparing
charged defects in to bulk into comparison of neutral defect to
bulk and further comparison of charged defect to neutral case.
Then, by explicitly writing the total energies for
a neutral defect, a charged defect in a small supercell, and
a charged defect in a large supercell, it is possible to inspect
the approximations hidden in each method. This framework should
also prove helpful in future development of methods for more
advanced methods and for more complex geometries.

We found that the method introduced by Freysoldt \cite{Freysoldt09}
works generally well.
Moreover, during the inspection of the potential alignment method,
we found a method which also worked surprisingly well, even if
formally looks as just a scaled Madelung-energy.

\begin{acknowledgments}
This work was supported by 
Tekniikan edist{\"{a}}miss{\"{a}}{\"{a}}ti{\"{o}}.
We are thankful for the Centre for Scientific Computing (CSC)
and Material Sciences National Grid Infrastructure (M-grid, Akaatti)
for the computational resources.
\end{acknowledgments}

%\newpage %Just because of unusual number of tables stacked at end
\bibliography{decor}% Produces the bibliography via BibTeX.

\end{document}